\documentclass[10pt]{article}
\usepackage{ametsoc2col}
\usepackage{subfigure}
\usepackage{float}
\usepackage{xcolor}
\usepackage{epstopdf}
\newcommand{\myabstract}{Numerical simulations of atmospheric circulation models are limited by their finite spatial resolution, and so large eddy simulation (LES) is the preferred approach to study these models. In LES a low-pass filter is applied to the flow field to separate the large and small scale motions. In implicitly filtered LES the computational mesh and discretization schemes are considered to be the low-pass filter while in the explicitly filtered LES approach the filtering procedure is separated from the grid and discretization operators and allows for better control of the numerical errors. The aim of this paper is to study and compare implicitly filtered and explicitly filtered LES of atmospheric circulation models in spectral space. To achieve this goal we present the results of implicitly filtered and explicitly filtered LES of a barotropic atmosphere circulation model on the sphere in spectral space and compare them with the results obtained from direct numerical simulation (DNS). Our study shows that although in the computation of some integral quantities like total kinetic energy and total enstrophy the results obtained from implicitly filtered LES and explicitly filtered LES show the same good agreement with the DNS results, explicit filtering produces energy spectra that show better agreement with the DNS results. More importantly, explicit filtering captures the location of coherent structures over time while implicit filtering does not.}

\begin{document}
%
%
\title{\textbf{\large{{Large Eddy Simulation of Turbulent  Barotropic Flows in Spectral Space on the Sphere}}}}
%
%
\author{\textsc{Leila N. Azadani,}
				\thanks{\textit{Corresponding author address:} 
				Leila N. Azadani,  Department of Engineering Science and Mechanics, Virginia Tech, Blacksburg, VA 24061, USA. 
				\newline{E-mail: leila@vt.edu}}\quad\textsc{and Anne E. Staples}\\
\textit{\footnotesize{ Department of Engineering Science and Mechanics, Virginia Tech, Blacksburg, VA 24061, USA}}
\and 
}
%
\ifthenelse{\boolean{dc}}
{
\twocolumn[
\begin{@twocolumnfalse}
\amstitle

\begin{center}
\begin{minipage}{13.0cm}
\begin{abstract}
	\myabstract
	\newline
	\begin{center}
		\rule{38mm}{0.2mm}
	\end{center}
\end{abstract}
\end{minipage}
\end{center}
\end{@twocolumnfalse}
]
}
{
\amstitle
\begin{abstract}
\myabstract
\end{abstract}
\newpage
}
%

\section{Introduction}
\label{sec:intro}
Atmospheric and oceanic general circulation models are key components of global climate models. The full form of the general circulation models are computationally expensive to be solved numerically. Therefore, different approximations are employed to simplify the full models and allow for detailed investigation of some specific effects. The barotropic vorticity equation (BVE) represents the simplest nontrivial model of the atmosphere which describes the evolution of a two-dimensional, nondivergent flow on the surface of the sphere. The BVE contains the nonlinear interactions of atmospheric motions and has been used extensively in the study of large-scale atmospheric dynamics. \cite{Charney} performed the first successful numerical weather prediction based on the BVE.

Atmospheric flows have a wide range of time and length scales which can vary from seconds to decades and from micrometers to several kilometers. Due to limited computational resources, resolving all of these scales numerically is not feasible in numerical simulations. Large eddy simulation (LES), in which the large scale motions are resolved explicitly and the effects of small scale motions are modeled, is the preferred method to solve these kinds of flow. 

In LES, a low-pass filter is applied to separate the flow field into large and small scale motions. The filtering operation in LES can be implicit or explicit. In implicit filtering the computational grid and discretization schemes are considered to be the low-pass filter that divides the flow field into resolved scale (RS) and subgrid scale (SGS) motions. In explicit filtering the filtering procedure is separate from the grid and discretization operations. The explicit filter width is usually larger than the grid spacing (implicit filter width), so in explicit filtering the flow field is divided into three portions, RS, resolvable subfilter scale (RSFS) and unresolvable subfilter scale (USFS) motions. Resolved scales are scales larger than the explicit filter width, whose contributions are computed numerically. Resolvable subfilter scales are those with a size between the explicit filter width and the implicit filter width. These scales can be reconstructed theoretically by using an inverse filtering operation. Unresolvable subfilter scales are scales smaller than the grid spacing and are known as subgrid scales, whose effects are typically modeled using an eddy viscosity model (\cite{Carati} and \cite{Zhou}). 

Implicit filtering is the most commonly used technique in LES of turbulent flows because it is computationally less expensive and less complicated than explicit filtering. However, implicit filtering is associated with some numerical issues (\cite{Lund}). Explicit filtering overcomes some of the difficulties associated with the implicit filtering and therefore has received increasing attention over the last few years. The accuracy of an explicitly filtered LES depends on three key factors, the filtering operation, the reconstruction model and the subgrid scale model. 

The filtering operation is performed by convolving a flow variable with the filter kernel. The most commonly used filter functions in LES of turbulent flows are the sharp cutoff filter, the Gaussian filter and the top-hat filter. If the filter width is constant, differentiation and filtering operations commute. Otherwise, a commutation error arises. \cite{Vasyliev} have developed a general theory for constructing continuous and discrete filters that commute with the differentiation up to any desired order in the filter width.  

The RSFS motions can be recovered by applying a reconstruction model. In physical space, reconstruction models are typically based on series expansion methods. The first reconstruction model was proposed by \cite{Leonard} where he provided an analytical expression based on Taylor series expansions of the filtering operator to reconstruct the filtered scales due to explicit filtering. The method was then improved by \cite{Clark} and is known as the gradient or nonlinear or tensor-diffusivity model. \cite{Bardina} presented the scale similarity model which assumes that the smallest resolved scales are similar to the largest unresolved scales. Thus, the unknown unfiltered quantities can be approximated by the filtered quantities. The velocity estimation model was proposed by \cite{Domaradzki}. In this model the unfiltered velocity field is estimated by expanding the resolved velocity field to subgrid scales two times smaller than the grid scale. The approximate deconvolution model (ADM) of \cite{Stolz1} is the most popular method for reconstructing the resolvable subfilter scales. In this model the the unfiltered flow quantities are approximated based on repeated application of an inverse filter to the filtered quantities. In spectral space, RSFS motions can be exactly recovered by convolving the inverse filter kernel with the filtered flow field. 

In contrast to reconstruction models, which are based solely on mathematical approximations and can be applied to both two- and three- dimensional turbulence, SGS models are developed based on the physical phenomenology of the problem. Since the behavior of two-dimensional turbulence is completely different from three-dimensional turbulence, the SGS models developed for LES of three-dimensional turbulence cannot be used in LES of two-dimensional turbulence. The first subgrid scale model for two-dimensional turbulence was probably developed by \cite{Leith} where he derived an eddy viscosity parameterization of the effects of unresolved scales on resolved scales using the eddy-damped Markovian approximation. \cite{Leith} applied the model to simulation of isotropic homogeneous large-scale atmospheric turbulence in Cartesian coordinates and found a good agreement between experimental and numerical results. \cite{Kraichnan} used the direct-interaction approximation (DIA) to develop an eddy viscosity subgrid scale model for two- and three-dimensional turbulence. Unlike the eddy-damped Markovian approximation, the DIA applies to inhomogeneous and anisotropic flows as well as homogeneous and isotropic flows. \cite{Boer} used the method proposed by  \cite{Leith} to develop a SGS model for computing large-scale atmospheric flows on the sphere in spectral space. \cite{Koshyk} modeled the nonlinear interactions between the resolved and unresolved scales in the simulation of general circulation models by using an empirical interaction function (EIF). The EIF function they used in their simulations was obtained based on a high-resolution computation. \cite{Frederiksen1} derived eddy viscosity and stochastic backscatter parameterizations for simulation of a two-dimensional atmospheric circulation model on the sphere. They used both eddy damped quasi-normal Markovian (EDQNM) and DIA closures in spherical geometry to derive some equations for eddy viscosity and stochastic backscatter for the forced-dissipative BVE. \cite{Frederiksen2} then improved the dynamic version of  Frederiksen and Daviesr's model by including the time-history effects in the stochastic modeling. \cite{Gelb} developed a spectral eddy viscosity model for computing shallow water flows in spherical geometry. Their model is based not on physical arguments, but rather on a mathematical approach to the problems exhibited at large wavenumbers in truncated spectral methods. 

Explicit filtering in LES of geophysical flows has been studied by a few researchers (\cite{Chow}, \cite{San1}, \cite{San2}). \cite{Chow} used ADM to reconstruct the RSFS motions and the dynamic Smagorinsky  model (\cite{Germano}) to parameterize the effects of SGS motions in the computations of the atmospheric boundary layer.  They found significant improvements in the accuracy of the results obtained from explicit filtering over the results obtained from implicit filtering. \cite{San1} and \cite{San2} used the ADM for recovering the RSFS term but did not used any SGS models. Results obtained using explicit filtering and ADM showed the correct four-gyre circulation structure predicted by direct numerical simulation (DNS) results in simulation of the forced BVE while the results obtained from implicit filtering yielded a two-gyre structure, which is not consistent with the DNS data. 

\cite{Azadani} have recently investigated the effects of explicit filtering on LES of turbulent barotropic flows in spectral space without applying any SGS model. Following \cite{Azadani} we would like to study the effects of explicit filtering on LES of two-dimensional atmospheric flows in spectral space by including the effects of SGS motions. We use a spectral method based on spherical harmonic transforms to solve the BVE in spectral space. A differential filter is applied to separate the flow field into resolved and subfilter scales. In order to reconstruct the unfiltered flow variables we use exact deconvolution by applying the exact inverse filter to the filtered flow field. The effects of subgrid scale term are taken into account by applying a spectral eddy viscosity term. 

The organization of the rest of this paper is as follows. The governing equations are presented in Section~\ref{sec:gov}. In Section~\ref{sec:nm} the numerical method is discussed. Reconstruction procedure and the LES equations are explained in Section~\ref{sec:adm}. The SGS model is presented is Section~\ref{sec:sgs}. The results and discussion are given in Section~\ref{sec:res} and conclusions are made in Section~\ref{sec:con}.

\section{Governing equations}
\label{sec:gov}
The BVE describes the motion of a two-dimensional, nondivergent, incompressible fluid on the rotating sphere and is given by
\begin{equation}
\frac{\partial\zeta}{\partial t}+J(\psi, \zeta+f)=\nu\nabla^{2}\zeta
\label{eq:bve}
\end{equation}
\begin{equation}
\zeta=\nabla^2\psi
\label{eq:laplas}
\end{equation}
where $\zeta(\lambda, \mu,t)$ is the vertical component of the vorticity, $\psi(\lambda, \mu, t)$ is the streamfunction, $f=2\Omega sin\theta$ is the Coriolis parameter, $\nu$ is the coefficient of the dissipation and $J$ is the horizontal Jacobian operator on the sphere, and is defined as
\begin{equation}
J(\psi, \zeta+f)=\frac{1}{R^2}\left [\frac{\partial\psi}{\partial\lambda}\frac{\partial(\zeta+f)}{\partial\mu}-\frac{\partial\psi}{\partial\mu}\frac{\partial(\zeta+f)}{\partial\lambda}\right]
\end{equation}
where $R$ is radius of the sphere, $-\pi\leqslant \lambda \leqslant \pi$ and $-\pi/2\leqslant \theta \leqslant \pi/2$ are the longitude and latitude, $\mu=sin\theta$, and $\Omega$ is the rotation rate of the sphere.

We nondimensionalize Eqs.~(\ref{eq:bve}) and (\ref{eq:laplas}) by using the radius of the sphere, $R$, as the length scale, $U$ as the characteristic velocity scale and $R/U$ as the advection time scale. Nondimensionalizing Eq.~(\ref{eq:bve}) introduces the Rossby number, an important physical parameter in a rotating system, which is defined as
\begin{equation}
Ro=\frac{U}{2R\Omega}
\label{eq:ro}
\end{equation}
and is a measure of the ratio of inertial forces to Coriolis forces. The Reynolds number, $Re=\frac{RU}{\nu}$, also appears as the coefficient of the dissipation term. 

The BVE is solved under periodic boundary conditions in the $\lambda$ direction. In the $\mu$ direction $\zeta$ should be independent of $\mu$ on the poles so
$$\zeta(\lambda, \mu, t)=\zeta(\lambda+2\pi, \mu, t) $$
and
$$ \zeta(\lambda, -1,t) \text{ and } \zeta(\lambda, 1, t) \text{ independent of } \lambda.$$

The initial conditions we use are based on the following initial energy spectrum (\cite{Cho})
\begin{equation}
E(n,0)=\frac {An^{\gamma/2}}{(n+n_0)^\gamma}
\label{eq:e0}
\end{equation}
where $A$ is a normalization constant, $n_0$ is the peak wavenumber of the energy spectrum, and $\gamma$ is used to control the width of the spectrum. Here, $n_0$ and $\gamma$ are set to 40 and 20, respectively.

\section{Numerical method}
\label{sec:nm}
The nondimensionalized  forms of  Eqs.~(\ref{eq:bve}) and (\ref{eq:laplas}) are solved in spectral space by expanding each variable into a series of spherical harmonics as
\begin{equation}
\zeta(\lambda, \mu, t)=\sum_{m=-\mathcal{N}}^{\mathcal{N}}\sum_{n=|m|}^{\mathcal{N}}\zeta_{n}^{m}(t)Y_{n}^{m}(\lambda, \mu)
\label{eq:sp}
\end{equation}
where $\zeta_{n}^{m}(t)$ are the complex spectral coefficients of $\zeta(\lambda, \mu, t)$, $n$ is the total wavenumber, $m$ is the zonal wavenumber, $\mathcal{N}$ denotes the truncation wavenumber and $Y_{n}^{m}$ are spherical harmonics defined by
\begin{equation}
Y_{n}^{m}(\lambda, \mu)=P_{n}^{m}(\mu)e^{im\lambda}
\label{eq:sh}
\end{equation}
where $P_{n}^{m}(\mu)$ are the normalized associated Legendre polynomials and $i=\sqrt{-1}$. Eq.~(\ref{eq:sp}) is truncated using triangular truncation which, unlike rhomboidal truncation, is rotationally symmetric. After nondimensionalizing and substituting the spectral representatives of $\zeta$ and $\psi$ into Eqs.~(\ref{eq:bve}) and (\ref{eq:laplas}), the BVE in spectral space is given by
\begin{equation}
\frac{\partial\zeta_n^m}{\partial t}=-J(\widehat{\psi_n^m,\zeta_n^m})-\frac{1}{Ro}im\psi_n^m-\frac{1}{Re}\frac{n(n+1)}{{R^*}^2}\zeta_n^m
\label{eq:bves}
\end{equation}
\begin{equation}
\zeta_n^m=-\frac{n(n+1)}{{R^*}^2}\psi_n^m
\label{eq:laplass}
\end{equation}
where $R^*=1$ and $J(\widehat{\psi_n^m,\zeta_n^m})$ is the nonlinear Jacobian term which is computed using the pseudospectral method.

The energy and enstrophy spectra are defined as
\begin{equation}
E(n,t)=\frac{1}{2}\sum_{m=-n}^{n}\frac{{R^*}^2}{n(n+1)}|\zeta_{n}^{m}(t)|^2
\label{eq:es}
\end{equation}
\begin{equation}
Ens(n,t)=\sum_{m=-n}^{n}n(n+1)E(n,t)
\label{eq:ens}
\end{equation}
and the total kinetic energy and enstrophy are given by
\begin{equation}
E(t)=\sum_{n=0}^{\mathcal{N}}E(n,t)
\label{eq:es}
\end{equation}
\begin{equation}
Ens(t)=\sum_{n=0}^{\mathcal{N}}Ens(n,t).
\label{eq:ens}
\end{equation}

\section{Exact deconvolution and the LES equations}
\label{sec:adm}
In LES, a spatial filter is applied to the fluid field to divide flow motions into large and small scales. The filtering operation is mathematically represented as a convolution product
\begin{equation}
\bar \zeta=G\ast \zeta
\label{eq:conv}
\end{equation}
where $\bar \zeta$ is the filtered variable, $G$ is the filter kernel, and the subfilter flow variable, $\zeta'$, is defined as
\begin{equation}
\zeta'=(1-G)\ast \zeta.
\label{eq:convu}
\end{equation}
In spectral space the convolution is turned into a multiplication
\begin{equation}
\bar \zeta_n^m=\hat G\zeta_n^m
\label{eq:convu}
\end{equation}
where $\hat G$ is the filter kernel in spectral space.

Applying the implicit filter due to the discretization scheme, along with an explicit filter to Eqs.~(\ref{eq:bves}) and (\ref{eq:laplass}), the governing equations of the LES are obtained as
\begin{eqnarray}
\frac{\partial \bar{\tilde{\zeta}}_n^m}{\partial t}&=&-J(\widehat{\bar{\tilde\psi}_n^m,\bar{\tilde\zeta}_n^m})-\frac{1}{Ro}im\bar{\tilde\psi}_n^m-\frac{1}{Re}\frac{n(n+1)}{{R^*}^2}\bar{\tilde\zeta}_n^m\nonumber\\
&&+{\tau_{n,}^m}_{SFS}
\label{eq:bvee}
\end{eqnarray}
\begin{equation}
\bar{\tilde\zeta}_n^m=\frac{n(n+1)}{{R^*}^2}\bar{\tilde\psi}_n^m
\label{eq:laplase}
\end{equation}
where the tilde indicates implicit filtering and the bar indicates explicit filtering. ${\tau_{n,}^m}_{SFS}$ is the subfilter scale stress and is defined as
\begin{eqnarray}
{\tau_{n,}^m}_{SFS}&=&J(\widehat{\bar{\tilde\psi}_n^m, \bar{\tilde\zeta}_n^m})-\widehat{\overline {J(\psi_n^m,\zeta_n^m})} \nonumber \\
&=&{\tau_{n,}^m}_{SGS}+ {\tau_{n,}^m}_{RSFS}
\label{eq:sfs}
\end{eqnarray}
where
\begin{eqnarray}
{\tau_{n,}^m}_{SGS}=\widehat{\overline{J(\tilde \psi_n^m,\tilde \zeta_n^m )}}-\widehat{\overline {J(\psi_n^m,\zeta_n^m)}}
\label{eq:sgs}
\end{eqnarray}
is the SGS stress to be addressed in Section~\ref{sec:sgs} and
\begin{eqnarray}
{\tau_{n,}^m}_{RSFS}=J\widehat{(\bar{\tilde\psi}_n^m, \bar{\tilde\zeta}_n^m)}-\widehat{\overline{J(\tilde \psi_n^m,\tilde \zeta_n^m)} }
\label{eq:rsfs}
\end{eqnarray}
is the RSFS stress. 

In spectral space, the unfiltered flow fields can be exactly reconstructed by applying the exact inverse filter to the filtered fields so, considering the effect of implicit filtering, the exact deconvolution of filtered flow variables is given by
\begin{equation}
\tilde \zeta_n^m=\bar{\tilde \zeta}_n^m/\hat G,
\label{eq:efz}
\end{equation}
\begin{equation}
\tilde\psi_n^m=\bar{\tilde\psi}_n^m/\hat G.
\label{eq:efp}
\end{equation}
The resolvable subfilter scale stress is then defined as
\begin{equation}
{\tau_{n,}^m}_{RSFS}=J(\bar{\tilde\psi}_n^m, \bar{\tilde\zeta}_n^m)-\overline{J(\bar{\tilde \psi}_n^m/\hat G,\bar{\tilde \zeta}_n^m/\hat G)}.
\label{eq:laplase}
\end{equation}

To perform explicit filtering in LES of the BVE we need to define an appropriate filter. We use the differential filter, which in physical space is defined as
\begin{equation}
\bar \zeta-\frac{\partial}{\partial x_j}\left(\alpha \frac{\partial \bar \zeta}{\partial x_i}\right)=\zeta
\label{eq:dif}
\end{equation}
where $\alpha=\alpha(x)$ is related to the filter width. We use the differential filter with a constant width, so Eq.~(\ref{eq:dif}) becomes
\begin{equation}
\bar \zeta-\alpha\nabla^2\bar \zeta=\zeta.
\label{eq:difc}
\end{equation}
Applying the definition of the Laplacian in spectral space
\begin{equation}
\nabla^2\zeta_{n}^{m}=-\frac{n(n+1)}{{R^*}^2}\zeta_{n}^{m}
\label{eq:lap}
\end{equation}
and substituting this into Eq.~(\ref{eq:difc}), the differential filter kernel in spectral space is defined as
\begin{equation}
\hat G(n)=\frac{1}{1+\alpha n(n+1)}.
\label{eq:sf}
\end{equation}
We use $\alpha =0.005$ in our computations.

\section{Subgrid scale model}
\label{sec:sgs}
SGS models have an important impact on the accuracy of LES results.  Inspired by the work has been done by \cite{Gelb} on shallow water computations, we use the following equation to model the SGS stress
\begin{equation}
{\tau_{n,}^m}_{SGS}=\epsilon{q}_n\zeta_{n}^{m}
\label{eq:nu}
\end{equation}
where 
\begin{eqnarray}
{q}_n =
\Bigg\{ \begin{array}{lr} 0  \\ exp\left[-\frac{\left(n-\mathcal{N}\right)^2}{\left(n-n_c\right)^2}\right] \end{array} \begin{array}{c} \quad   \quad n \leq n_c  \\ \quad  \quad n_c < n \leq\mathcal{N}, \end{array}
\label{eq:qn}
\end{eqnarray}
$\epsilon$ is the spectral eddy viscosity amplitude, and $n_c$ is the cutoff wavenumber. $\epsilon$ and $n_c$ depend on the truncation wavenumber, $\mathcal N$. To overcome the failure of convergence \cite{Maday} suggested the following dependencies of $\epsilon$ and $n_c$ on $\mathcal N$
\begin{equation}
\epsilon\sim\frac{1}{\mathcal N},\hspace{1cm}         n_c\sim 5\sqrt{\mathcal N},
\label{eq:en}
\end{equation}
We found that the $\epsilon$ obtained from the above relation is too large and makes the model developed here too dissipative. Comparison of the LES results obtained with the above SGS model with the results obtained from DNS showed that the following relations for $\epsilon$ and $n_c$ can produce LES results that agree better with DNS results in our computations.

\begin{equation}
\epsilon\sim\frac{1}{\mathcal N^2},\hspace{1cm}         n_c\sim \mathcal N^{\frac{3}{4}}.
\label{eq:en}
\end{equation}

\begin{figure*}[!t]
\begin{center}
\subfigure[]{
\resizebox*{5.3cm}{!}{\includegraphics[trim = 20mm 10mm 20mm 20mm, clip, width=3cm]{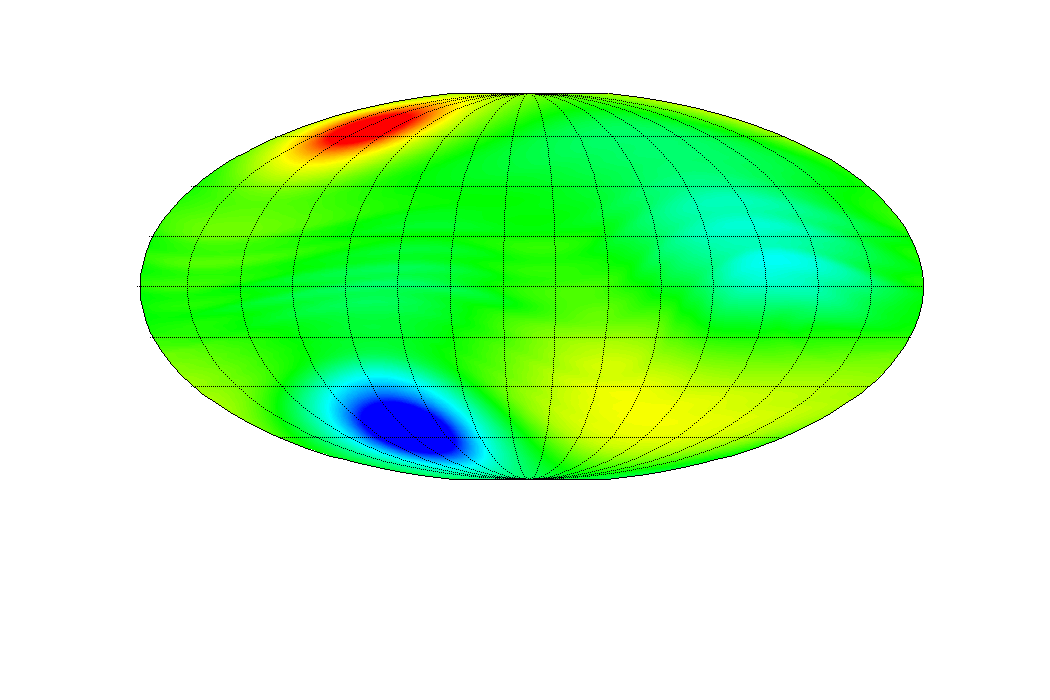}}}%
\subfigure[]{
\resizebox*{5.3cm}{!}{\includegraphics[trim = 20mm 10mm 20mm 20mm, clip, width=3cm]{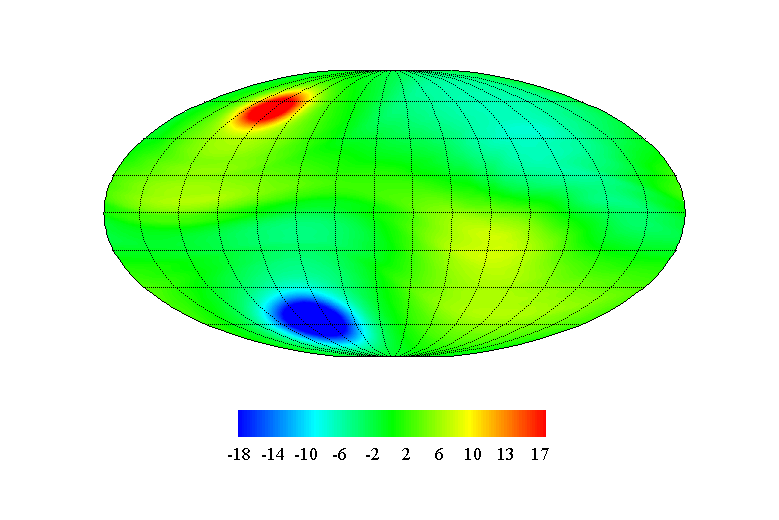}}}%
\subfigure[]{
\resizebox*{5.3cm}{!}{\includegraphics[trim = 20mm 10mm 20mm 20mm, clip, width=3cm]{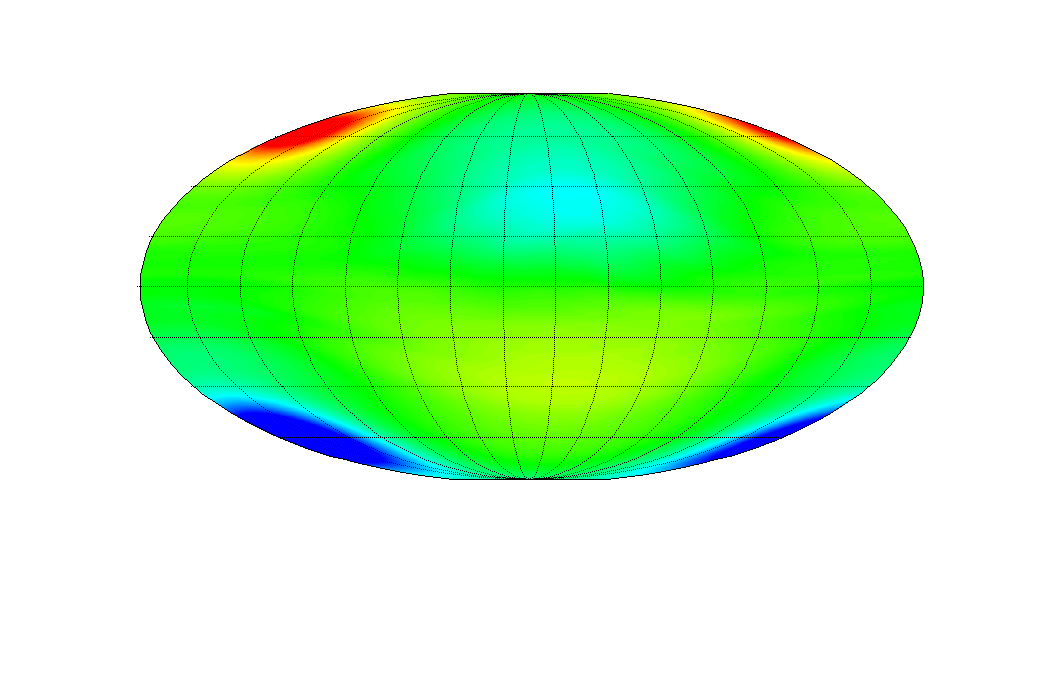}}}%
\caption{The vorticity field for the (a) DNS (resolution T333), (b) explicitly filtered LES (resolution T66), and (c) implicitly filtered LES (resolution T66) results for Experiment 1 ($Re=10^{4}$, $Ro=0.05$).}
\label{fig:v0}
\end{center}
\end{figure*}

\begin{figure*}[!t]
\begin{center}
\subfigure[]{
\resizebox*{5.3cm}{!}{\includegraphics[trim = 20mm 10mm 20mm 20mm, clip, width=3cm]{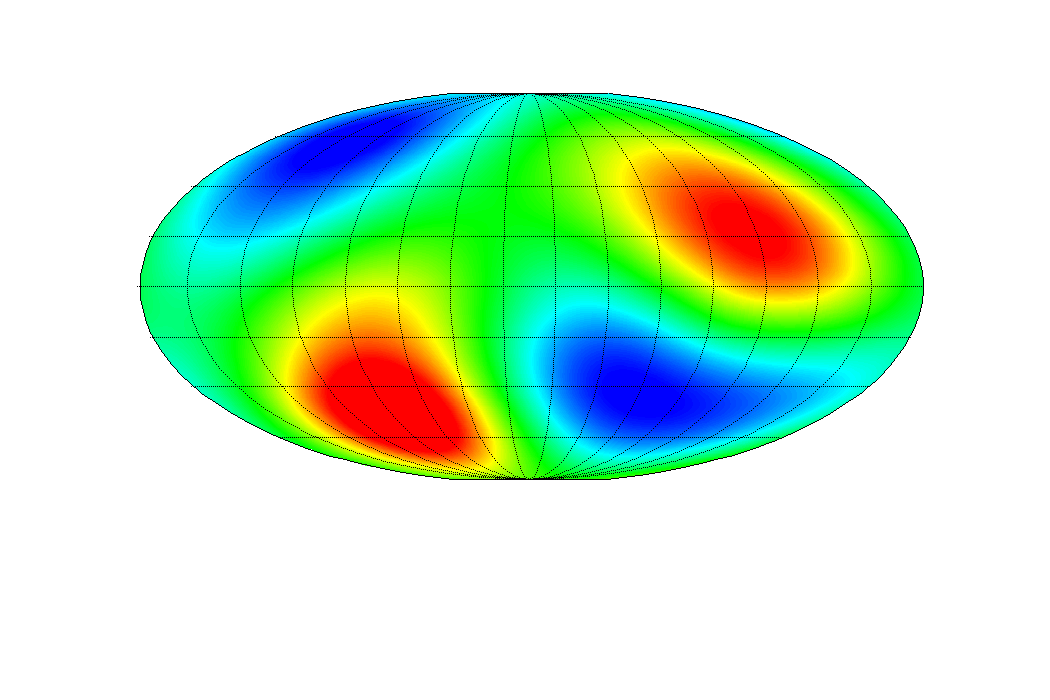}}}%
\subfigure[]{
\resizebox*{5.3cm}{!}{\includegraphics[trim = 20mm 10mm 20mm 20mm, clip, width=3cm]{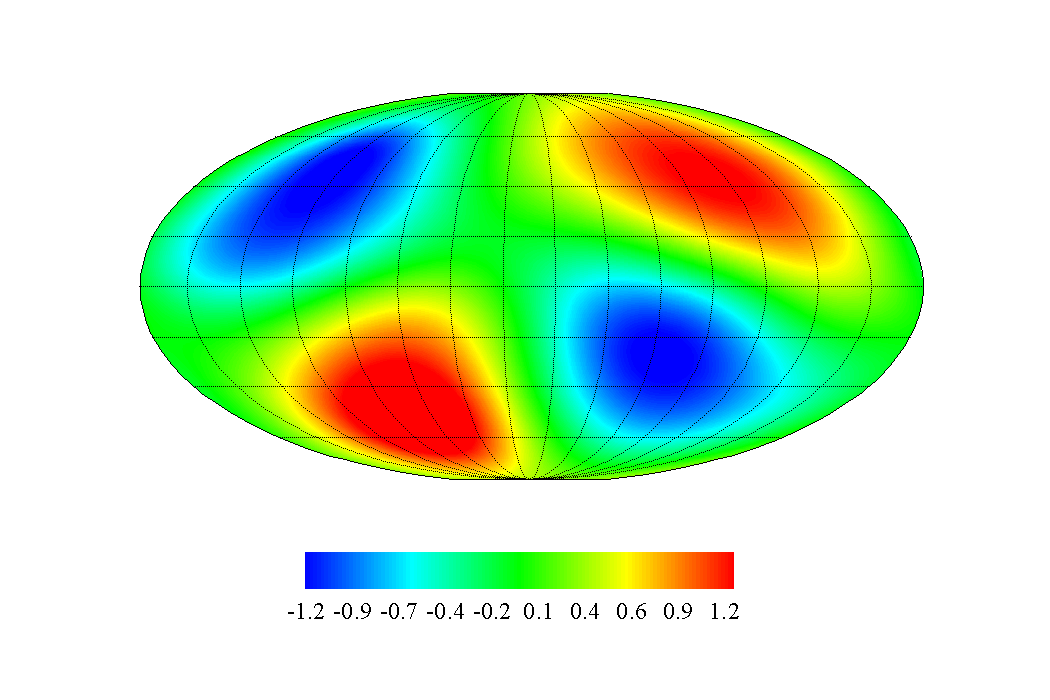}}}%
\subfigure[]{
\resizebox*{5.3cm}{!}{\includegraphics[trim = 20mm 10mm 20mm 20mm, clip, width=3cm]{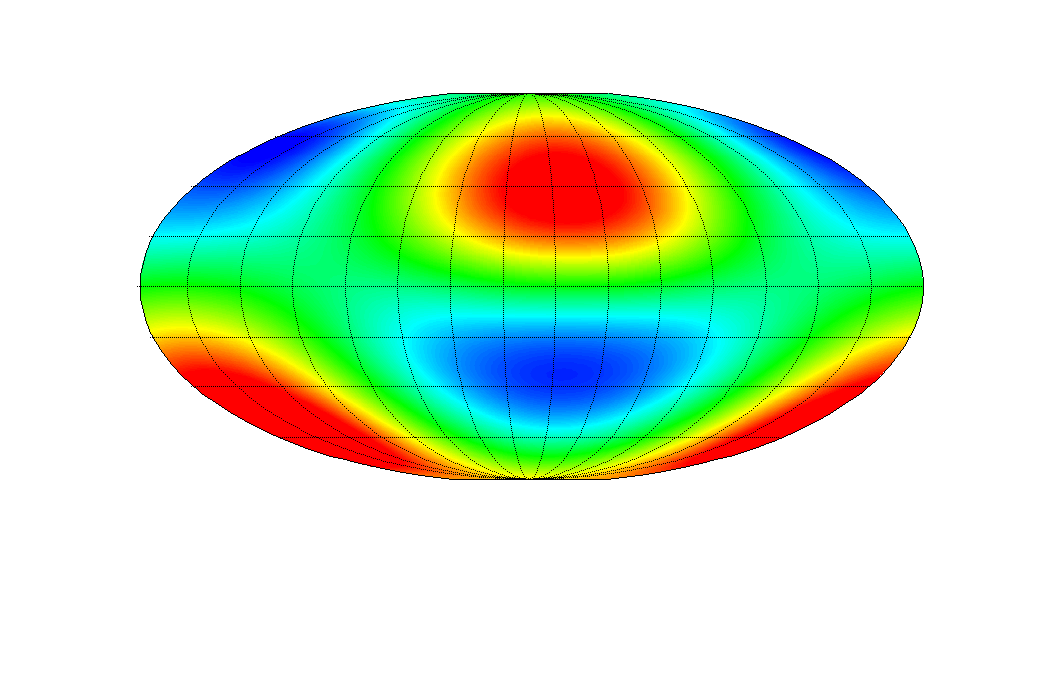}}}%
\caption{The streamfunction field for the (a) DNS (resolution T333), (b) explicitly filtered LES (resolution T66), and (c) implicitly filtered LES (resolution T66) results for Experiment 1 ($Re=10^{4}$, $Ro=0.05$).}
\label{fig:p0}
\end{center}
\end{figure*}

\section{Results and discussion}
\label{sec:res}
Two numerical experiments are performed. One experiment simulates the behavior of a barotropic fluid at a small Rossby number where the rotation of the sphere is large and the other experiment solves the BVE at a large Rossby number when the rotation is small. In both experiments the Reynolds number is $Re=10^{4}$. The resolution of the DNS computations is $T=333$ which corresponds to $1000$(longitude)$\times500$(latitude) grid points in physical space. The resolution in the implicitly and explicitly filtered LES runs is $T66$ or $200$(longitude)$\times100$(latitude). In this paper, we refer to DNS as the computation with high resolution in which the coherent structures are properly resolved, it does not signify direct numerical simulation in which all the turbulent structures are resolved numerically. The fourth order Runge-Kutta scheme is used for the time integration of the governing equations and the 2/3 dealiasing rule is applied to compute the nonlinear term.
\subsection{Experiment 1}
The first experiment is performed at $Ro=0.05$. Contour plots of the vorticity and streamfunction for the DNS, explicitly filtered LES and implicitly filtered LES are shown in Figs.~\ref{fig:v0} and \ref{fig:p0}, respectively. These plots show the turbulent coherent structures at $t=50$. It can be seen that the explicitly filtered LES data can predict the exact number and correct location of the positive and negative vortices while the implicitly filtered LES results do not track the correct behavior of the coherent structures. 

\begin{figure}[!t]
\begin{center}
\includegraphics[width=19pc]{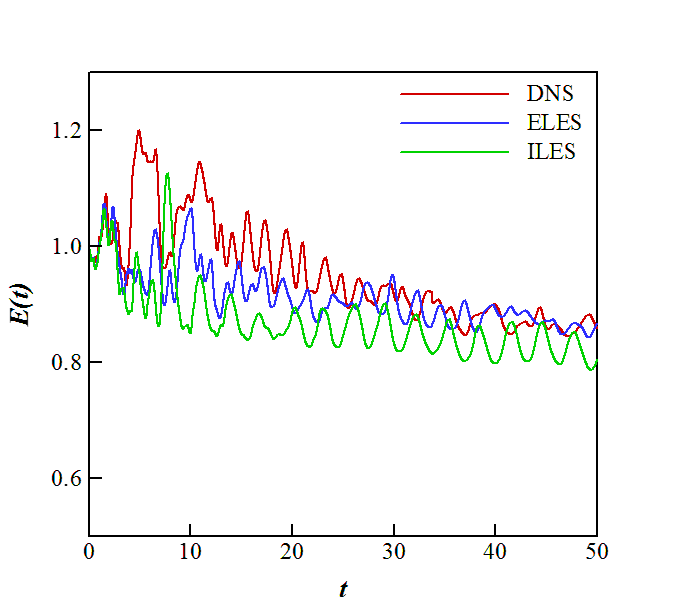}%
\caption{Comparison of the total kinetic energy for the DNS, implicitly filtered LES (ILES), and explicitly filtered LES (ELES) results for Experiment 1 ($Re=10^{4}$, $Ro=0.05$).}
\label{fig:ke0}
\end{center}
\end{figure}

The variation of the total kinetic energy with time is presented in Fig.~\ref{fig:ke0}. This figure shows that although at early times both implicitly filtered and explicitly filtered LES results do not show good agreement with the DNS results, at larger times the results obtained from the explicitly filtered LES follow the DNS data more closely than the results from the implicitly filtered LES.

\begin{figure}[!b]
\begin{center}
\includegraphics[width=19pc]{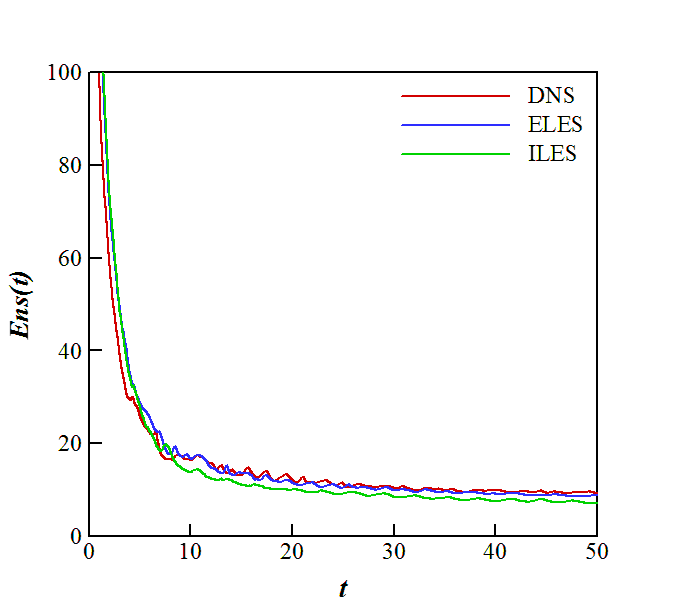}%
\caption{Decay of the total enstrophy for the DNS, implicitly filtered LES (ILES), and explicitly filtered LES (ELES) results for Experiment 1 ($Re=10^{4}$, $Ro=0.05$).}
\label{fig:ens0}
\end{center}
\end{figure}

Fig.~\ref{fig:ens0} shows the temporal variation of the total enstrophy. It can be seen that the results obtained from the explicitly filtered LES are at the same level as the DNS results while the results obtained from the implicitly filtered LES seem to be too dissipative and predict a lower level for the total enstrophy.

\begin{figure}[!t]
\begin{center}
\includegraphics[width=19pc]{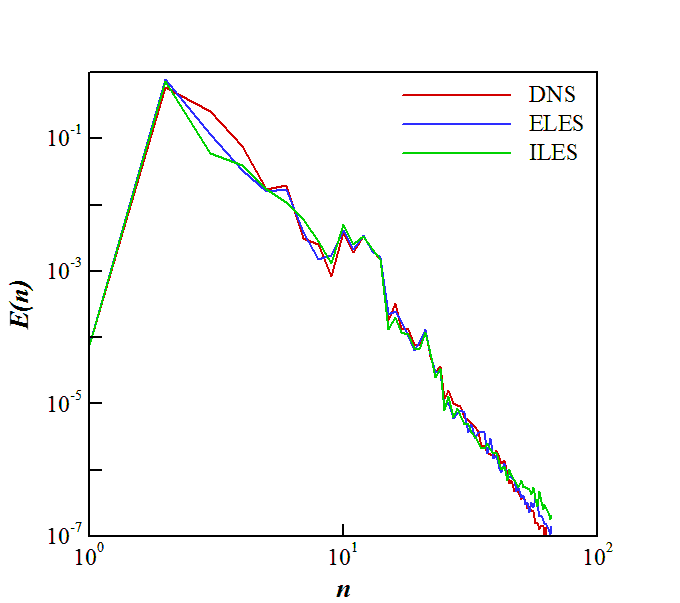}%
\caption{Variation of the energy spectrum with wavenumber for the DNS, implicitly filtered LES (ILES), and explicitly filtered LES (ELES) results for Experiment 1 ($Re=10^{4}$, $Ro=0.05$).}
\label{fig:es0}
\end{center}
\end{figure}

The variation of the energy spectrum with wavenumber is shown in Fig.~\ref{fig:es0}. This plot is made at $t=20$, when the flow has passed the transient state and become fully developed. Both implicitly and explicitly filtered LES perform very well in the computation of energy spectrum however the explicitly filtered LES results are in better agreement with the DNS results, particularly at the highest wave numbers.

\subsection{Experiment 2}
\begin{figure*}[!t]
\begin{center}
\subfigure[]{
\resizebox*{5.3cm}{!}{\includegraphics[trim = 20mm 10mm 20mm 20mm, clip, width=3cm]{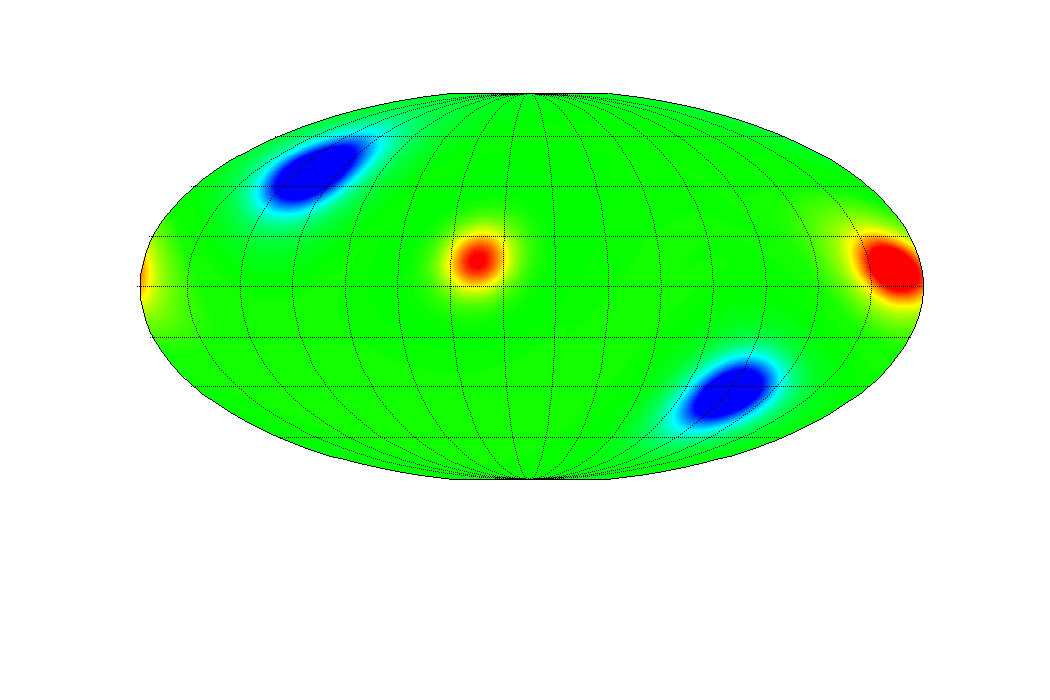}}}%
\subfigure[]{
\resizebox*{5.3cm}{!}{\includegraphics[trim = 20mm 10mm 20mm 20mm, clip, width=3cm]{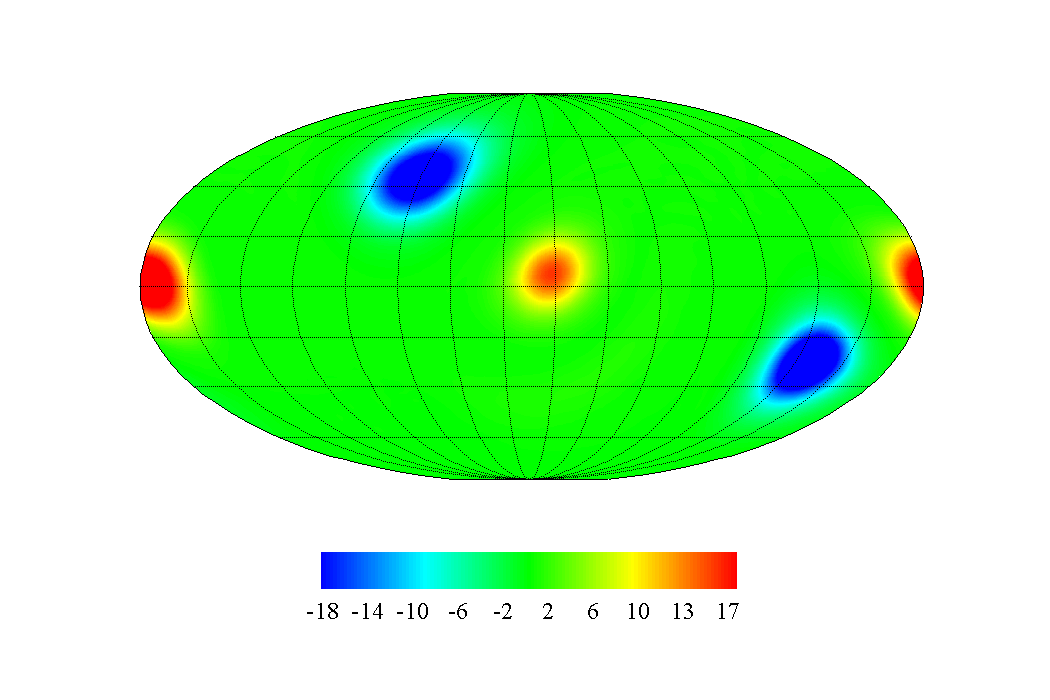}}}%
\subfigure[]{
\resizebox*{5.3cm}{!}{\includegraphics[trim = 20mm 10mm 20mm 20mm, clip, width=3cm]{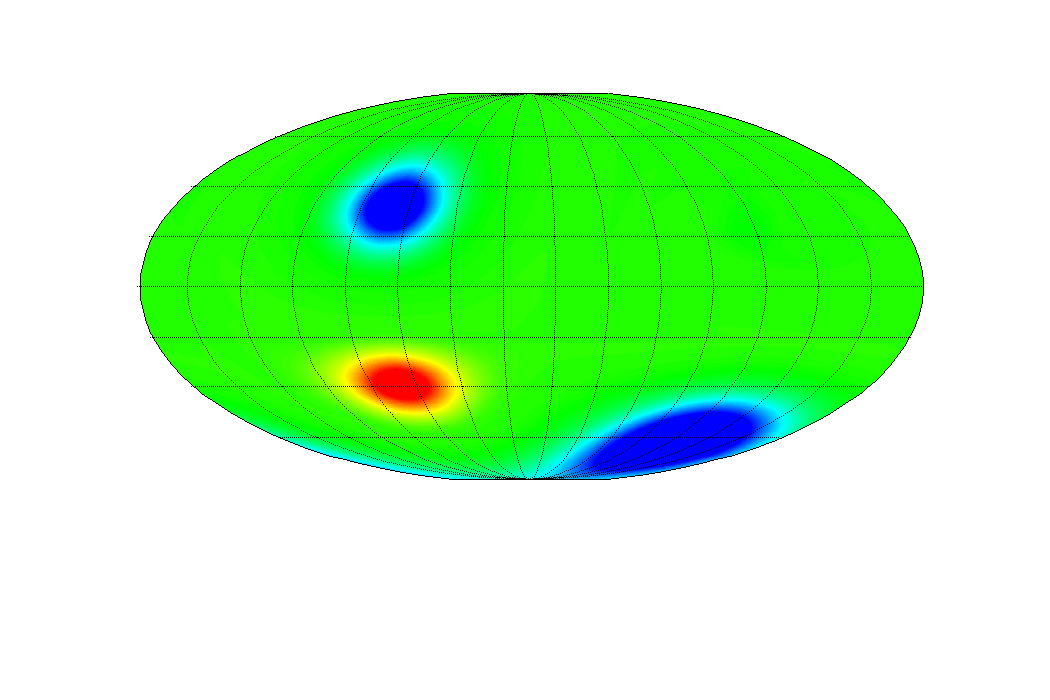}}}%
\caption{The vorticity field for the (a) DNS (resolution T333), (b) explicitly filtered LES (resolution T66), and (c) implicitly filtered LES (resolution T66) results for Experiment 2 ($Re=10^{4}$, $Ro=5$).}
\label{fig:v1}
\end{center}
\end{figure*}

\begin{figure*}[!t]
\begin{center}
\subfigure[]{
\resizebox*{5.3cm}{!}{\includegraphics[trim = 20mm 10mm 20mm 20mm, clip, width=3cm]{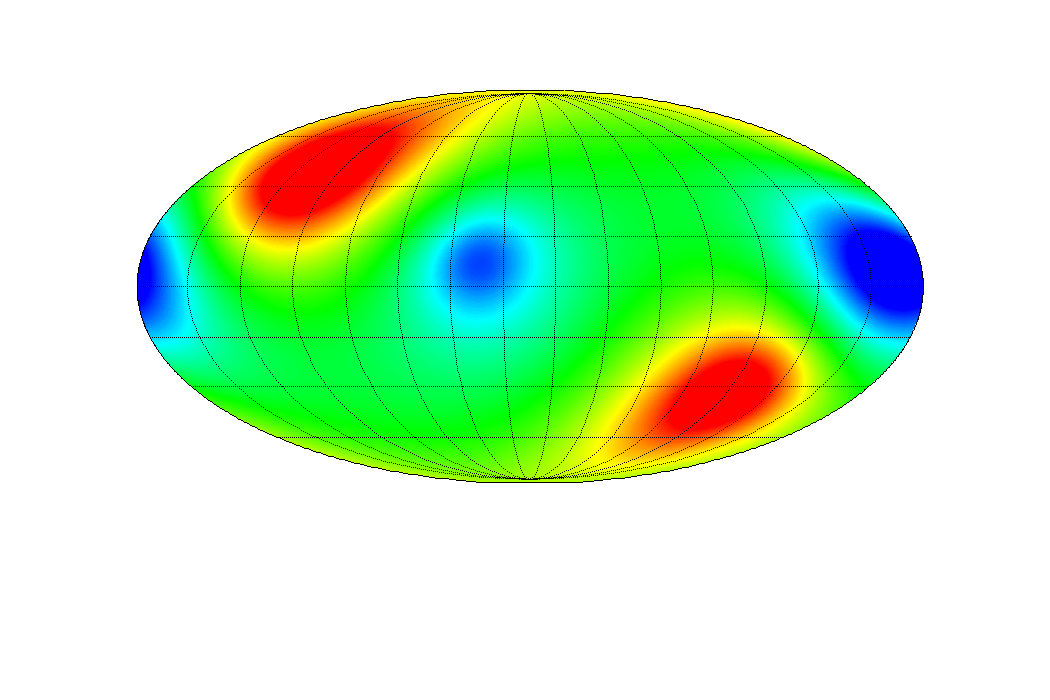}}}%
\subfigure[]{
\resizebox*{5.3cm}{!}{\includegraphics[trim = 20mm 10mm 20mm 20mm, clip, width=3cm]{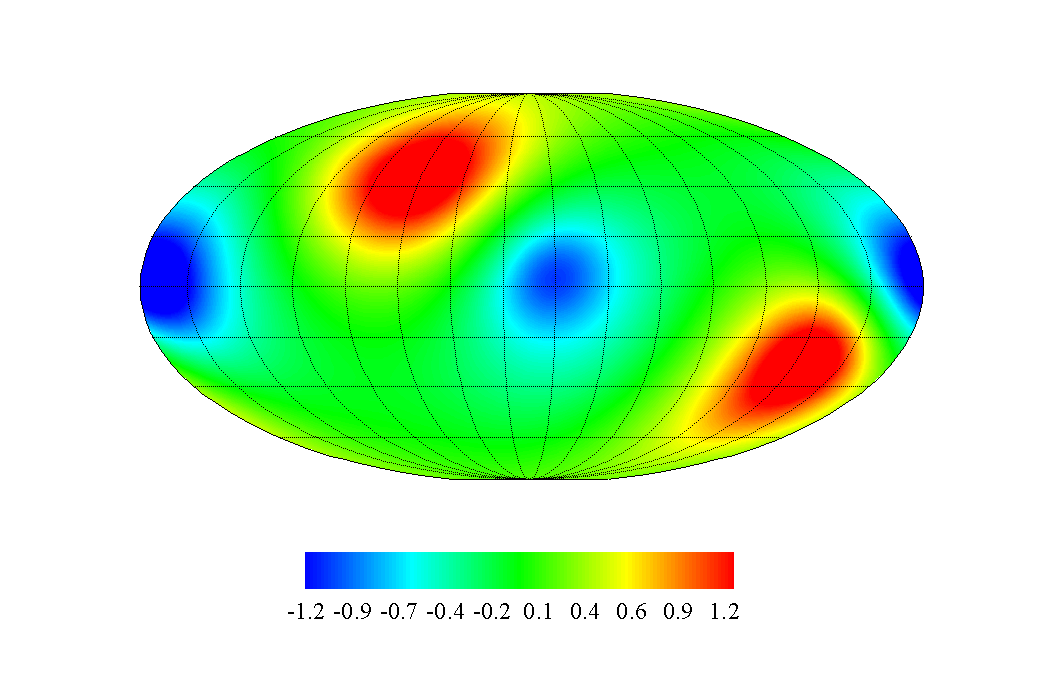}}}%
\subfigure[]{
\resizebox*{5.3cm}{!}{\includegraphics[trim = 20mm 10mm 20mm 20mm, clip, width=3cm]{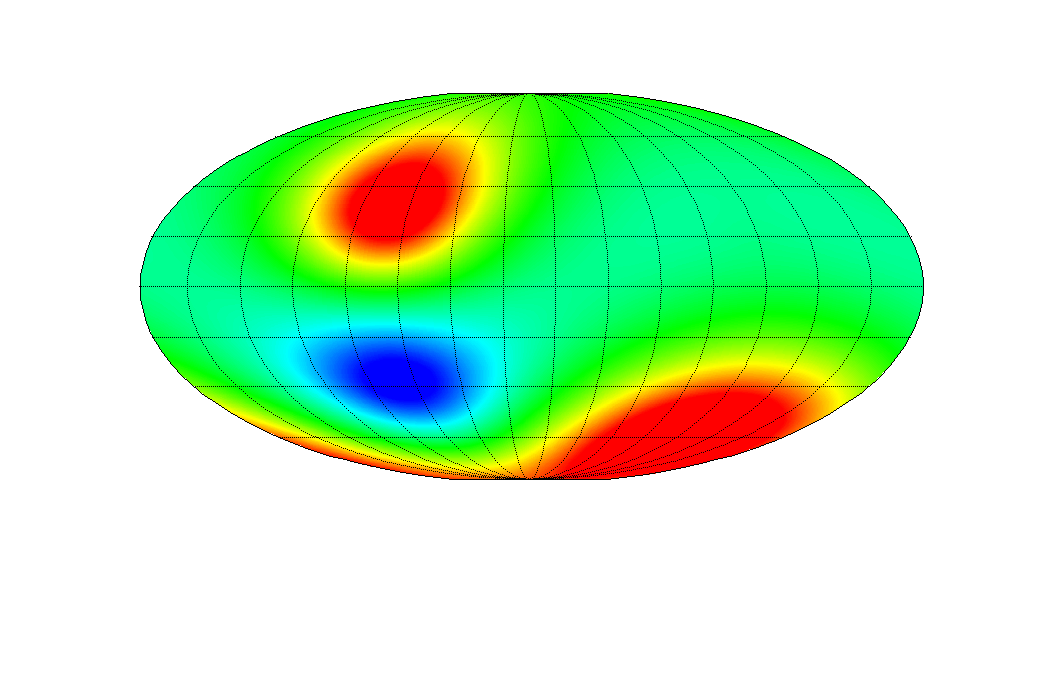}}}%
\caption{The stream function field for the (a) DNS (resolution T333), (b) explicitly filtered LES (resolution T66), and (c) implicitly filtered LES (resolution T66) results for Experiment 2 ($Re=10^{4}$, $Ro=5$).}
\label{fig:p1}
\end{center}
\end{figure*}

Experiment 2 is performed at $Ro=5$. Vorticity and streamfunction fields are shown in Figs.~\ref{fig:v1} and \ref{fig:p1}, respectively. Long-time integration of the BVE at high Rossby numbers produces a vortical quadrupole state (\cite{Cho}) which can be seen in Figs.~\ref{fig:v1} and \ref{fig:p1}. Although Figs.~\ref{fig:v0} and \ref{fig:p0} show the vortical quadrupole state too, it is not always the case at small Rossby numbers.  As in Experiment 1 it can be seen that the results obtained from the explicitly filtered LES are a nearly perfect match with the DNS data, while the implicitly filtered LES results show an incorrect vortical tripole state.
 
\begin{figure}[!b]
\begin{center}
\includegraphics[width=19pc]{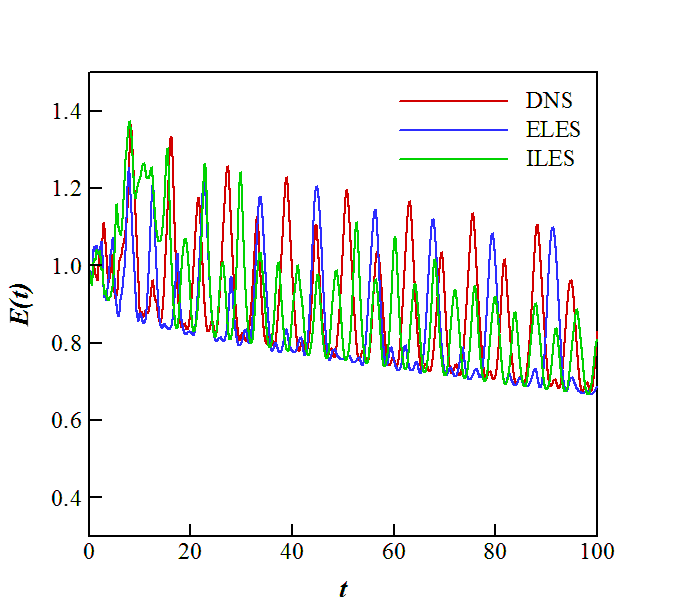}%
\caption{Comparison of the total kinetic energy for the DNS, implicitly filtered LES (ILES), and explicitly filtered LES (ELES) results for Experiment 2 ($Re=10^{4}$, $Ro=5$).}
\label{fig:ke1}
\end{center}
\end{figure}

\begin{figure}[!b]
\begin{center}
\includegraphics[width=19pc]{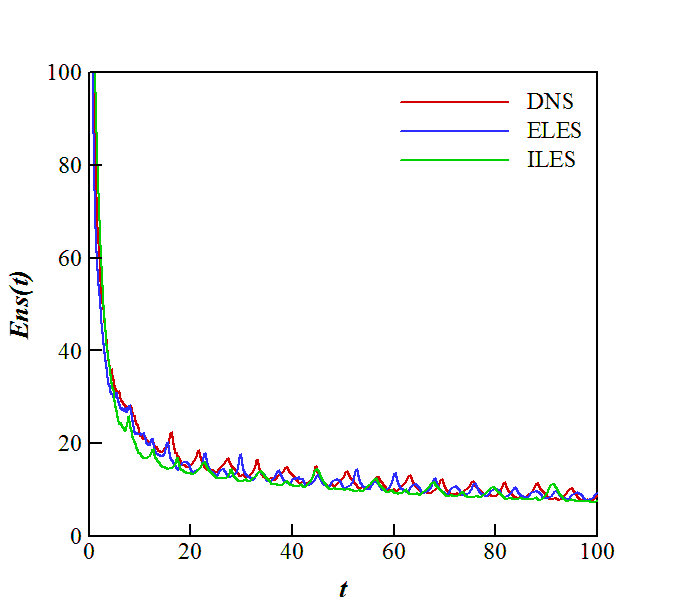}%
\caption{Decay of the total enstrophy for the DNS, implicitly filtered LES (ILES), and explicitly filtered LES (ELES) results for Experiment 2 ($Re=10^{4}$, $Ro=5$).}
\label{fig:ens1}
\end{center}
\end{figure}

The temporal variation of the total kinetic energy is shown in Fig.~\ref{fig:ke1}. The results obtained from implicitly and explicitly filtered LES are both at the same level as the DNS results and show the correct behavior of the total kinetic energy. 

The variation of the total enstrophy with time is shown in Fig.~\ref{fig:ens1}. Similar to Fig.~\ref{fig:ens0} the results obtained from the implicitly and explicitly filtered LES both show good agreement with the DNS data, although here the implicitly filtered LES performs a bit better than for low Rossby number results in Experiment 1.

\begin{figure}[!t]
\begin{center}
\includegraphics[width=19pc]{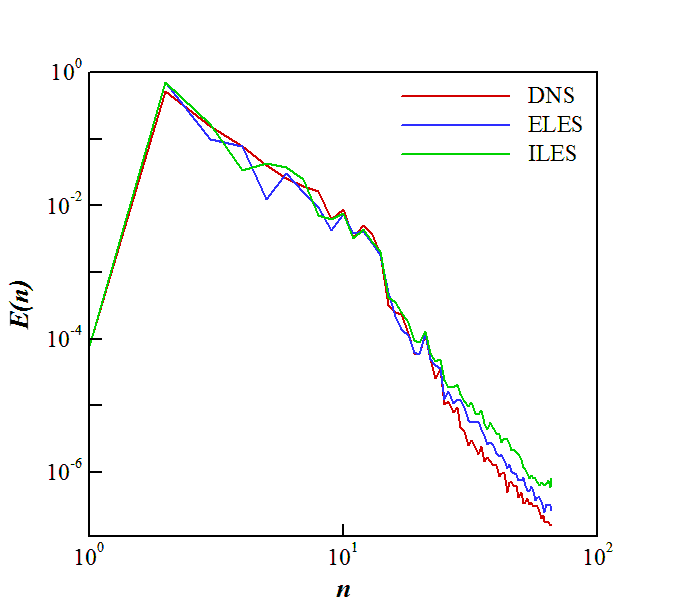}%
\caption{Variation of the energy spectrum with wavenumber for the DNS, implicitly filtered LES (ILES), and explicitly filtered LES (ELES) results for Experiment 2 ($Re=10^{4}$, $Ro=5$).}
\label{fig:es1}
\end{center}
\end{figure}

Fig.~\ref{fig:es1} shows the decay of the energy spectrum with wavenumber at $t=20$ when the flow is fully developed. It can be seen that the explicitly filtered LES results show even better agreement with the DNS data at high wavenumbers than they did in Experiment 1.

In all comparisons, the explicitly filtered LES results show a better agreement with the DNS results in capturing the behavior of the coherent structures. Although in the computation of some quantities the results obtained from the implicitly filtered LES appear to be as accurate as the results obtained from the explicitly filtered LES, in other cases the explicitly filtered LES performs better than the implicitly filtered LES.  
 
\section{Conclusions}
\label{sec:con}
We studied large eddy simulation (LES) of the turbulent barotropic vorticity equation (BVE) on the sphere in spectral space. Both implicitly and explicitly filtered LES were investigated and results obtained from implicit filtering were compared with the results obtained from explicit filtering. In explicitly filtered LES, the differential filter was used to separate large and small scales and exact deconvolution was applied to reconstruct the unfiltered flow variables from the filtered flow variables. A spectral eddy viscosity model was used to parameterize the effects of subgrid scales on resolved scales in both implicit and explicit filtering computations. We performed two sets of numerical experiment, one at a small Rossby number and the other one at a large Rossby number. Comparison of the results obtained for the total kinetic energy, total enstrophy and energy spectrum from both experiments showed that although in some cases implicit filtering and explicit filtering give similarly accurate results, the explicit filtering results agree better overall with the results obtained from direct numerical simulation (DNS). The superior performance of explicit filtering over implicit filtering was seen in the contour plots of the vorticity and streamfunction. While explicitly filtered LES results  predicted the exact location of the coherent structures, implicit filtering was not able to capture the the correct behavior of the coherent structures, even predicted the wrong number of coherent structures in some cases. Coherent structures play an important role in the computation of atmospheric flows and need to be captured correctly for accurate numerical weather prediction and climate modeling. The results obtained in this paper show that applying an explicit filter in addition to an implicit filter in LES of a turbulent barotropic flow in spectral space can improve the results and accurately predict the behavior of the coherent structures in the flow.

\ifthenelse{\boolean{dc}}
{}
{\clearpage}
\bibliographystyle{ametsoc}
\bibliography{references}

\end{document}